 \documentclass[pmlr,twocolumn,10pt]{jmlr} 

\usepackage{multirow}
\usepackage{booktabs}
\usepackage{enumerate}
\usepackage{siunitx}
\usepackage{xcolor,colortbl}
\usepackage[switch]{lineno}
\usepackage{ulem}

\title[The Role of Graph-based MIL and Interventional Training in the Generalization of WSI Classifiers]{The Role of Graph-based MIL and Interventional Training in the Generalization of WSI Classifiers}

\author{\Name{Rita Pereira}\Email{rita.martins.pereira@tecnico.ulisboa.pt}\\
\AND
\Name{Mª Rita Verdelho}\Email{ritaverdelho@hotmail.com}\\
\AND
\Name{Catarina Barata} \Email{ana.c.fidalgo.barata@tecnico.ulisboa.pt}\\
\AND
\Name{Carlos Santiago}
\Email{carlos.santiago@tecnico.ulisboa.pt} \\
\addr Institute for Systems and Robotics, LARSyS, Instituto Superior Técnico, University of Lisbon, Lisbon, Portugal
}

\begin{document}

\maketitle

\begin{abstract}
Whole Slide Imaging (WSI), which involves high-resolution digital scans of pathology slides, has become the gold standard for cancer diagnosis, but its gigapixel resolution and the scarcity of annotated datasets present challenges for deep learning models. Multiple Instance Learning (MIL), a widely-used weakly supervised approach, bypasses the need for patch-level annotations. However, conventional MIL methods overlook the spatial relationships between patches, which are crucial for tasks such as cancer grading and diagnosis. To address this, graph-based approaches have gained prominence by incorporating spatial information through node connections. Despite their potential, both MIL and graph-based models are vulnerable to learning spurious associations, like color variations in WSIs, affecting their robustness. In this dissertation, we conduct an extensive comparison of multiple graph construction techniques, MIL models, graph-MIL approaches, and interventional training, introducing a new framework, \textbf{G}raph-based \textbf{M}ultiple \textbf{I}nstance \textbf{L}earning with \textbf{I}nterventional \textbf{T}raining (GMIL-IT), for WSI classification. We evaluate their impact on model generalization through domain shift analysis and demonstrate that graph-based models alone achieve the generalization initially anticipated from interventional training. Our code is available here: \href{https://github.com/ritamartinspereira/GMIL-IT}{github.com/ritamartinspereira/GMIL-IT}.
\end{abstract}

\section{Introduction}
\label{sec:intro}


In computational pathology, analysis of Whole Slide Images (WSI) is the gold standard for cancer diagnosis \citep{Zhang2022}. WSIs are high-resolution digital representations of entire pathology slides \citep{HOQUE2022, Haan2021}. 
The typical approach for WSI analysis consists of pathologists manually annotating each Region of Interest (ROI) with a microscope. A slide contains multiple ROIs, making the process time-consuming and largely contributing to the limited availability of annotated datasets for WSI \citep{CAD2022}. Recently, there has been a notable increase in research on machine learning approaches to assist pathologists in disease classification, screening, tissue localization, and distinguishing benign from malignant regions \citep{CAD2022}. 


Multiple Instance Learning (MIL) has become widely used for WSI analysis. In MIL, a WSI is divided into smaller patches, known as instances, and grouped into bags, with classification occurring at the bag level \citep{KOMURA2018, Li2021}. However, existing MIL methods often neglect the spatial relationship between patches, which is a critical factor for tasks like cancer grading or cancer classification \citep{Raju2020}. The importance of considering the spatial correlation between different patches becomes even more apparent when acknowledging that pathologists, during the manual annotation of WSI, are also aware of the correlations among entities throughout the slide rather than isolated areas. For this reason, graph-based approaches have gained attention for being able to model these relationships within WSIs \citep{Guan2022, Bontempo2023, das-milBontempo}.

Nonetheless, ensuring the above models are robust remains a challenge, as they are vulnerable to spurious associations (biases) that compromise their generalization. Such biases are caused by changes in imaging instruments, hospitals, and staining techniques, as well as air bubbles that appear during slide preparation or markers used by pathologists to highlight areas of interest \citep{Brixtel2022, devilindetails}.

We propose a novel evaluation setup based on Camelyon16 \citep{Bejnordi2017} and Camelyon17 \citep{Bandi2019, LitjensCam17}, where we explicitly enforce domain shifts. We conduct an extensive evaluation of several graph representation strategies and integrate them in popular MIL approaches. Additionally, we proposed a new Graph-based Multiple Instance Learning framework with Interventional Training (GMIL-IT), leveraging graphs to capture spatial context between patches and applying interventional training through backdoor adjustment. This evaluation aims to evaluate how each of these strategies enhance the model robustness to biases and domain shifts, contributing to more accurate diagnosis.

Our key contributions are summarized as follows:
\begin{enumerate}[(i)]
    \item Comparison of multiple graph construction methods for WSI representation.
    \item Analysis of the impact of interventions on graph-based MIL models. 
    \item Qualitative and quantitative analysis of model robustness in a domain shift.
\end{enumerate}
The analysis showed that graph-based models alone outperform models enhanced with interventional training, highlighting the robustness of the graph structures.

\section{Related Work}
\label{sec:related_work}
\subsection{Multiple Instance Learning}


MIL is a weakly supervised learning method that works with data having uncertain or ambiguous labels, or when only the slide-level label is available \citep{Fatima2023}.

Under this setting, the entire WSI is partitioned into multiple patches, and these form a bag. The goal is to train a model that maps from a group of instances, that compose a bag, to a single bag-level label \citep{Tarkhan2022}.


Several approaches have suggested different pooling functions to derive bag embeddings from instances. \citep{Ilse2018} proposed using an attention mechanism to generate attention scores for each instance. The bag embedding is then computed as the weighted average of the instance embeddings. \citep{Li2021} introduces a dual-stream architecture for generating bag embeddings. In the first stream, an instance classifier operates on instance-level embeddings, followed by max pooling to extract the highly scored feature representing the critical instance. In the second stream, a pooling operator is applied to aggregate the instance embeddings and generate the bag embedding. 



\subsection{Graph-based methods}
Several studies have applied graph-based methods to WSI analysis for tasks like survival analysis \citep{Li2018, WANG2024}, lymph node metastasis prediction \citep{Zhao2020} and cancer staging \citep{Raju2020, Shi2023}.

WSIs can be represented as graphs using different strategies for node definitions, including cell-graphs, tissue-graphs, and patch-graphs. \citep{Zhou2019CGCNetCG} introduced cell-graphs, where nodes represent cell nuclei and edges capture cellular interactions. Alternatively, \cite{Anklin2021_incomplete} proposed tissue-graphs, with nodes representing tissue superpixels and edges encoding interactions between tissue regions. However, both cell-graphs and tissue-graphs have limitations, cell-graphs miss tissue macro-architecture, while tissue-graphs overlook cellular interactions.
For this reason, patch-graphs are a good option because they provide a balance between fine-grained detail and computational efficiency, capturing important spatial relationships without the overhead of cell-level precision or the loss of detail seen in tissue-graphs.

\citep{Chen2021} proposed constructing a patch-graph convolutional network, where each node represents a patch in the WSI. The edges connect neighboring patches and are defined based on the spatial coordinates of the patches in the WSI, rather than feature similarity as some of the other works have explored \citep{Li2018, ding2024}. 

In our approach, we explore different graph construction methods, each with distinct edge and node definitions. We conduct a comparative study to evaluate their impact on both performance and representation accuracy.



\subsection{Causal Inference}

Causal inference is an important research topic with application in a variety of fields, including economics \citep{Hal2016}, social sciences \citep{Imbens2024}, machine learning \citep{Cui2020} and education \citep{Kim2018}. Its increased attention in recent years is due to its capacity to address and eliminate harmful confounding effects on model predictions. 
Recently, it has also drawn increasing attention in the medical domain, particularly in WSI analysis, where the aim is to reduce spurious associations caused by confounders for more accurate diagnoses. \citep{Lin2023} suggests applying causal interventions through backdoor adjustment to achieve deconfounded predictions. However, this method requires an observable set of confounders, which may not always be available. To address this limitation, \citep{Chen2024} introduces CaMIL, which utilizes frontdoor adjustment, eliminating the need to explicitly model the confounder set. Since modeling a confounder set is feasible in our case, we apply the backdoor adjustment in our model to produce unbiased predictions.

\section{Methods}
\label{sec:methods}
\begin{figure*}[htb]
\floatconts
  {fig:nodes}
  {\caption{General Pipeline of GMIL-IT. The process begins with a feature extractor and graph construction module \ref{subsec:FeatGraph}. A GNN model is then used to generate spatially aware instance representations, with a MIL aggregator creating the bag embedding and a classifier determining the label \ref{subsec:GraphMIL}. Bag-level clustering is performed to build a confounder dictionary, which is concatenated to the bag embeddings during the Interventional Training stage \ref{subsec:IT}.}}
  {\includegraphics[width=1.0\linewidth]{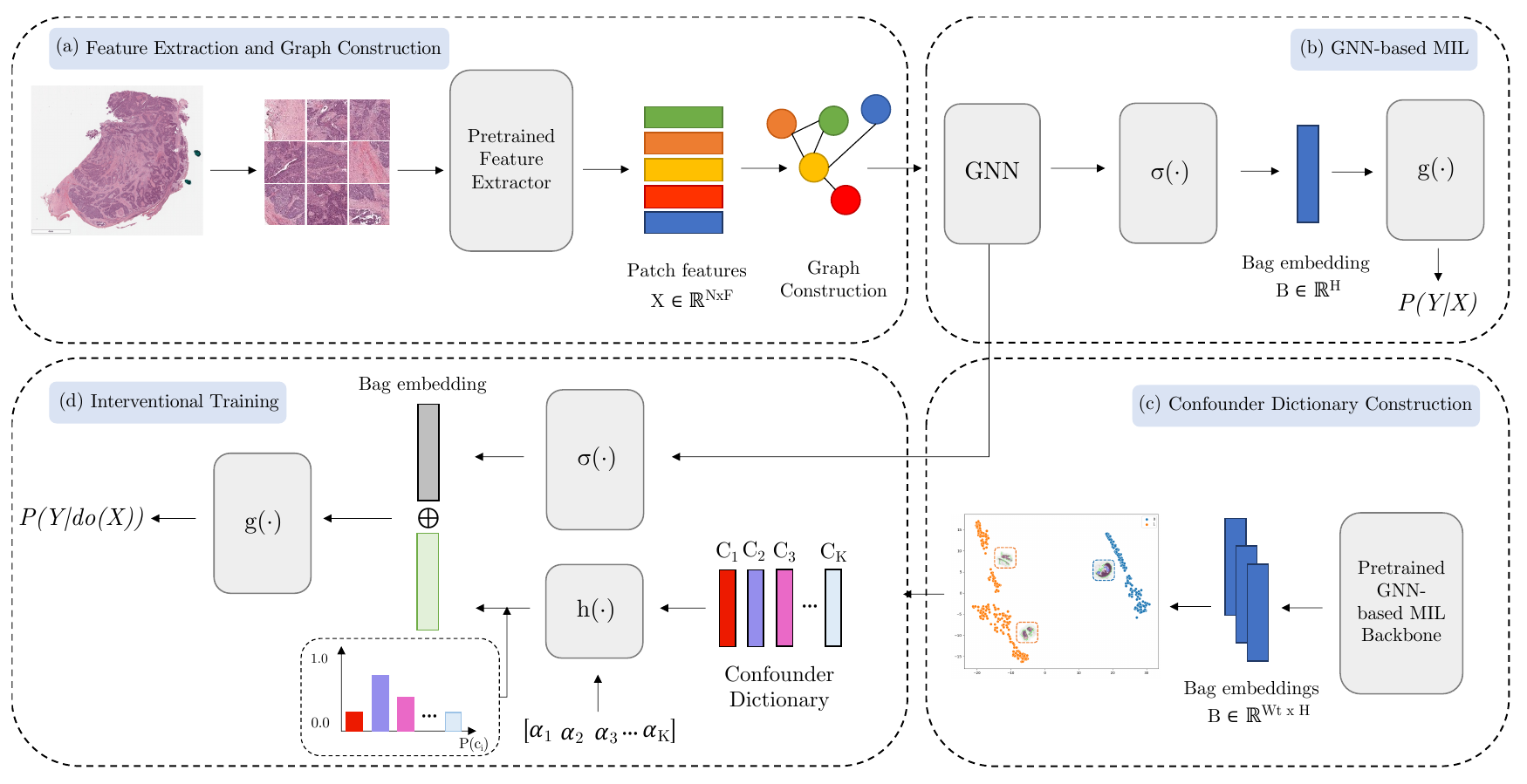}}
  \label{fig:architecture}
\end{figure*}

Figure \ref{fig:architecture} illustrates the pipeline comprising of three main components: feature extractor and graph construction module \ref{subsec:FeatGraph}, the GNN-based MIL pipeline \ref{subsec:GraphMIL} and finally the Interventional Training stage \ref{subsec:IT}. For each module, we compare several approaches, with the combination of all the components resulting in GMIL-IT.

\subsection{Feature Extraction and Graph Construction}
\label{subsec:FeatGraph}
To generate feature embeddings, we adopt the patching method and feature extractor outlined in \citep{Lu2021}.
Each WSI is first segmented to remove the background and any internal holes. The segmented image is then divided into non-overlapping patches at a 20x magnification, with each patch sized 256x256 pixels. These patches are input into a ResNet-50 pretrained on ImageNet, which converts them into 1024-dimensional feature embeddings. Each image is represented by a node feature matrix of size $X \in \mathrm{R}^{N_p \times F}$, where $N_p$ is the number of patches in the image, and $F$=1024. We opted to not use foundational models for feature extracting to prevent a potential data leakage.

In this work, we propose exploring various methods for representing the WSI as a graph, including approaches that operate globally on the entire dataset (Region-Graphs) and those applied locally to individual images (Patch-Graphs and Centroid-Graphs).

\paragraph{Patch-Graphs:} Each WSI is represented as a patch-graph, with each node corresponding to a patch and characterized by a feature vector. Unlike other patch-graph methods that define the adjacency matrix in the embedding space, we adopt the approach from \citep{Chen2021}, defining edges in the Euclidean space.
Using the previously extracted patch coordinates, we construct an adjacency matrix $A$ based on the spatial proximity of the patches. An edge is established between two nodes $i$ and $j$ as
\begin{equation}
A_{i,j} = \left\{\begin{matrix}
1 & \text{if} \; j \in \mathcal{N}(i),\\ 
0 & \text{otherwise}
\end{matrix}\right.
\end{equation}
where $\mathcal{N}(i)$ represents the set of nodes adjacent to node $i$. 

\paragraph{Region-Graphs:} To enable the model to cluster similar textures across the entire dataset, we experimented with applying K-means clustering. By considering the entire dataset, K-means can capture global patterns and relationships that might be missed when clustering images individually. 

Due to the high computational demands, processing all the images at once would be unfeasible. Therefore, we processed 50 images at a time and applied a partial fit using Mini Batch K-means \citep{Sculley2010}, which operates on small subsets of data in each iteration. 
Each patch from the WSI was assigned to the nearest centroid. To create a Region Adjacency Graph (RAG), regions were identified using Connected Component Analysis. Each node in the RAG represents a segmented region, with its feature vector being the average of the patch features in that region. Edges denote spatial adjacency between regions.

\paragraph{Centroid-Graphs:} Patch-graphs correspond to an extreme case where all the information from the WSI is kept. Centroid-graphs and Region-Graphs explore ways of representing the WSI in a compact manner.
For each WSI, we perform k-means clustering on the patch-level feature embeddings. This results in $k$ centroids, where each centroid represents a cluster of patches within the image. The centroids of each cluster are then used as nodes in the graph, with each node represented by a feature vector, defined as the mean of the feature vectors of all patches in the cluster. To construct the edges of the graph, we begin by defining a fully connected graph with an adjacency matrix $A$ initialized to ones. Next, we assign edge weights based on the cosine distance between the centroids' feature vectors. However, by following this approach, we lose the spatial information since we are not using the original patch coordinates.

\subsection{Graph-based MIL}
\label{subsec:GraphMIL}
The Graph-based MIL model (figure \ref{fig:architecture} (b)) combines GNN and a MIL aggregator to learn spatial relationships and produce the final representation.

\paragraph{Graph Neural Networks:} Following the graph construction module, each WSI is represented as $G\left( X, A\right) $, where $X \in \mathrm{R}^{N_p \times F}$ corresponds to the node feature matrix and $A$ defines the adjacency matrix.

The graph serves as input to a Graph Neural Network (GNN) that enable nodes to exchange information and to capture the spatial structure of the {WSI}. It consists of a sequence of $L$ GNN layers, where $L$ is an adjustable hyperparameter that will be explored later, that integrate information from the $L$-th order neighborhood. The initial node features, denoted as $X \in \mathbb{R}^{N_p \times F}$, are processed through multiple GNN layers, resulting in a tensor of shape $\mathbb{R}^{N_p \times D}$, where $D$ represents the size of the hidden channels. We experimented with two types of GNN: Graph Convolution Networks (GCN) and Graph Attention Networks (GAT). 

In the GCN proposed by \citep{kipf2017semisupervised}, the new node feature matrix $H$ after the $l$-th layer of GCN is given by
\begin{equation}
    H^{\left( l + 1 \right)} = \alpha\left( \tilde{D}^{-1/2} \tilde{A} \tilde{D}^{-1/2} H^{\left( l \right)} W^{\left( l \right)} \right),
    \label{eq:GCN}
\end{equation}
where $A$ is the adjacency matrix of the graph, $\tilde{A} = A + I_n$ is the adjacency matrix with self-connections, $\mathbf{\tilde{D}}$ is the diagonal node degree matrix of $\mathbf{\tilde{A}}$, $W$ is matrix of weights, and $H$ is matrix of activations, with $H^{\left( 0\right)} = X$.

GAT \citep{velickovic2018} employs a graph attention layer that assigns attention scores to the neighborhood of node $v$. This allows the model to assign different importance to nodes in the neighborhood, as well as provide explanability.

In the graph attention layer, a linear transformation is first applied to the whole graph, parametrized by a weight matrix $W$. Then, an attention mechanism $a$ is employed to derive attention coefficients $e_{vu} = a\left (\textbf{W}_{h_v}, \textbf{W}_{h_u} \right )$, indicating the significance of node $u's$ features for node $v$. These coefficients are calculated only for the nodes $u \in N_v$. The raw attention scores $e_{vu}$ are then normalized across all neighbors of node $u$ using the softmax function to obtain normalized attention coefficients $\alpha_{vu}$.

The new node features are computed by applying a weighted sum of the original node features with the previously calculated attention coefficients, followed by a sigmoid activation function,
\begin{equation}
\vec{h}_v^{'} = \sigma\left ( \sum_{u \in N_v} \alpha_{vu} \textbf{W}\vec{h}_u \right ).
    \label{eq:GAT_noderep}
\end{equation}
After $L$ GNN layers, we obtain the final node feature matrix $H^L$.

\paragraph{Multiple Instance Learning:}
In our approach, we use MIL to aggregate node-level features into a bag-level representation, which is then used for classification. 

A typical MIL approach is defined by a three-staged framework. It consists of a transformation $f$ applied to the individual instances, a permutation-invariant pooling function $\phi$ that aggregates these instance features to form a bag-level representation and a classifier $g$ to make the final prediction for the bag.
We experiment with two MIL pooling methods: ABMIL \citep{Ilse2018} and DSMIL \citep{Li2021}. The resulting embedding is passed through a classifier $g$ to obtain the prediction. This is formally defined as
\begin{equation}
    \hat{y}_{stg2} = g\left( \sigma\left( H^{L}\right) \right) = g(B).
\end{equation}

The loss for the GNN-based MIL model is given by binary cross-entropy
\begin{equation}
\begin{split}
      \mathcal{L}_{stg2} =  -\frac{1}{N}\sum _{i = 1}^N Y_i\log \left (\hat{Y_i}^{stg2}\right ) \\ + \left ( 1 - Y_i \right )\log\left ( 1 - \hat{Y_i}^{stg2} \right ).
  \label{eq:loss2}
\end{split}
\end{equation}

\subsection{Interventional Training}
\label{subsec:IT}

The Interventional Training stage consists of two main components: Confounder Dictionary and Backdoor Adjustment. Given that the causal graph for our problem is unknown, the interventional training framework used in this work follows the one described in IBMIL \citep{Lin2023}. Since backdoor adjustment is model-agnostic, it can be applied to our framework.


We define the backdoor adjustment formula as
\begin{equation}
    \begin{split}
        P\left( Y| do\left( X \right) \right) = P\left( Y|X \right)\sum_{i}P\left( c_i \right) \\ = \sum_{i} P\left( Y|X, c_i \right)P\left( c_i \right).
    \end{split}
    \label{eq:backdoor_initial}
\end{equation}

\paragraph{Confounder Dictionary:}
To implement \eqref{eq:backdoor_initial}, we first define the observable confounder set $C$. We begin by extracting bag features from the GNN-based MIL framework for the training set $B \in \mathrm{R}^{N_t \times F}$, where $N_t$ is the number of WSIs in the training set. The dimensionality of these bag embeddings is then reduced using PCA to facilitate clustering. Subsequently, K-means is applied to cluster the bag embeddings into $K$ clusters, which likely represent visual biases in the images (e.g., variations in color staining).

The confounder dictionary $C = \left[ c_1, c_2, \dots, c_K\right]$ consists of several confounder strata $c_i$, which are created by averaging the features of each of the $K$ clusters. Here, $K$ is a hyperparameter that determines the size of the confounder dictionary.

\paragraph{Backdoor Adjustment:}
Using equation \eqref{eq:backdoor_initial} as a reference, the backdoor adjustment formula for the problem at hand is defined by
\begin{equation}
    P\left( Y | do\left( X \right) \right) = \sum_{i} P\left( Y | X, h\left( X, c_i \right) \right) P\left( c_i \right).
    \label{eq:backdoor_ibmil}
\end{equation}
In this equation, $h\left(X,c_i \right)$ introduces an attention mechanism, that reflects the relevance of confounder $c_i$ to the bag embedding $B$, according to
\begin{equation}
    \begin{matrix}
h\left ( X, c_i \right ) = \alpha_i c_i,
\\  \;
\left [ \alpha_1, \dots, \alpha_K \right ] = \text{softmax}\left ( \frac{\left ( W_1B \right )^T \left ( W_2C \right )}{\sqrt{l}} \right ),
\end{matrix}
\label{eq:attentionconf}
\end{equation}
where $B$ is the bag feature vector $\phi(f(X))$, $C$ is the confounder dictionary, $W_1$ , $W_2$ are learnable projection matrices, $\alpha_i $ is the attention score for confounder $i$ and $l$ denotes vector normalization.

Finally, $P\left (Y|do\left (X\right )\right )$ is expressed as 
\begin{equation}
  P\left ( Y | do\left (X\right ) \right ) \approx P\left ( Y \Bigg| B\oplus \sum_{i=1}^K \alpha_ic_iP\left ( c_i \right ) \right ),
\label{eq:intervention}
\end{equation}
where $\oplus$ denotes vector concatenation , $\alpha_i$ corresponds to the attention scores and $P\left (  c_i \right )$ denotes the probability of confounder $c_i$. $B\oplus \sum_{i=1}^K \alpha_ic_iP\left ( c_i \right )$ is then fed to a feed-forward neural network.
This is an approximation given by the Normalized Weighted Geometric Mean \citep{Xu2015}, where we simplify the computation by shifting the outer sum into the Softmax function. The intervention combines the bag features $B$ with the function $h\left ( \cdot \right )$. Finally, the loss used for stage 3 is similar to \eqref{eq:loss2}.


\section{Experiments Setup}
\label{sec:experiments}
\subsection{Datasets}

\textbf{Camelyon16} \citep{Bejnordi2017} is a public dataset for metastasis detection in breast cancer. The dataset comprises 399 whole-slide images sourced from two medical centers in the Netherlands: Radboud University Medical Center (RUMC) and University Medical Center Utrecht (UMCU). The acquisition process involved two different scanning methods: RUMC utilized a scanner equipped with a 20x objective lens, while UMCU employed a digital slide scanner featuring a 40x objective lens.

\textbf{Camelyon17} \citep{Bandi2019, LitjensCam17}, similar to Camelyon16, is a public dataset for breast cancer. It comprises 1000 WSI collected from five medical centers in the Netherlands: RUMC, UMCU, the Rijnstate Hospital (RST), the Canisius-Wilhelmina Hospital (CWZ) and LabPON (LPON). For these experiments, we use 500 slides corresponding to 100 patients. Slides from RUMC, CWZ, and RST were digitized using the 3DHistech Pannoramic Flash II 250 scanner at RUMC. At UMCU, a Hamamatsu NanoZoomer-XR C12000-01 scanner was used, while at LPON, a Philips Ultrafast Scanner was employed for scanning.

\subsection{Implementation Details}
All experiments were conducted on a system equipped with NVIDIA GeForce RTX 3090 graphics card featuring 24 GB of GDDR6X RAM. We used 5-fold cross-validation for both datasets. For Camelyon16, we shuffled data from the two medical centers and then applied cross-validation. In contrast, for Camelyon17, each test fold contained data from a single medical center, highlighting the dataset's inherent domain shifts.

The experiments were done using the Adam optimizer with a learning rate of 1e-4 for the MIL model and 1e-3 for the GNN model. Weight decay was set to 1e-4 for the MIL model and 5e-4 for the GNN model. The models were trained for 50 epochs with a batch size of 1 and gradient accumulation of 8. We evaluated the classification performance using AUC, Balanced Accuracy, F1 score, and Precision, reporting the average across the 5 test sets. 

\subsection{Comparative Analysis Details}
We use state-of-the-art MIL models, including ABMIL \citep{Ilse2018} and DSMIL \citep{Li2021}, as baselines. For the comparison of graph construction methods, we explore different approaches to defining nodes and edges, such as Patch-Graphs, Region-Graphs, and Centroid-Graphs. Additionally, we experiment with various GNNs, including GCN \citep{kipf2017semisupervised} and GAT \citep{velickovic2018}. Both the graph-based models and MIL models are evaluated with and without interventional training.

\begin{table}[t]
\caption{Performance of various graph-based models on Camelyon16, obtained with $L=3$ and max pooling.}
\centering
\setlength{\tabcolsep}{4pt} 
\resizebox{\columnwidth}{!}{ 
\begin{tabular}{@{}clccc@{}}
\toprule
\multicolumn{2}{c}{\multirow{2}{*}{\textbf{Model}}} & \multicolumn{3}{c}{\textbf{Camelyon16}}            \\
\multicolumn{2}{c}{}                                & \textbf{Accuracy} & \textbf{Recall} & \textbf{AUC} \\ \midrule
\multicolumn{2}{c}{Patch-GCN} & 0.912 $\pm$ 0.02 & 0.900 $\pm$ 0.04 & 0.900 $\pm$ 0.04 \\ \midrule
\multicolumn{2}{c}{Patch-GAT} & \textbf{0.922 $\pm$ 0.03} & \textbf{0.908 $\pm$ 0.03} & \textbf{0.908 $\pm$ 0.03} \\ \midrule
\multicolumn{2}{c}{Global Region-GCN} & 0.732 $\pm$ 0.03 & 0.722 $\pm$ 0.03 & 0.722 $\pm$ 0.03 \\ \midrule
\multicolumn{2}{c}{Global Region-GAT} & 0.744 $\pm$ 0.05 & 0.745 $\pm$ 0.05 & 0.745 $\pm$ 0.05 \\ \midrule
\multicolumn{2}{c}{Local Region-GCN} & 0.774 $\pm$ 0.04 & 0.754 $\pm$ 0.04 & 0.754 $\pm$ 0.04 \\ \midrule
\multicolumn{2}{c}{Local Region-GAT} & 0.704 $\pm$ 0.06 & 0.703 $\pm$ 0.06 & 0.703 $\pm$ 0.06 \\ \midrule
\multicolumn{2}{c}{Centroid-GCN} & 0.709 $\pm$ 0.07 & 0.701 $\pm$ 0.07 & 0.701 $\pm$ 0.07 \\ \midrule
\multicolumn{2}{c}{Centroid-GAT} & 0.702 $\pm$ 0.09 & 0.691 $\pm$ 0.09 & 0.691 $\pm$ 0.09 \\ \bottomrule
\end{tabular}
}
\label{tab:graphs_representation}
\end{table}

\begin{figure*}[t]
  {\includegraphics[width=1.0\linewidth]{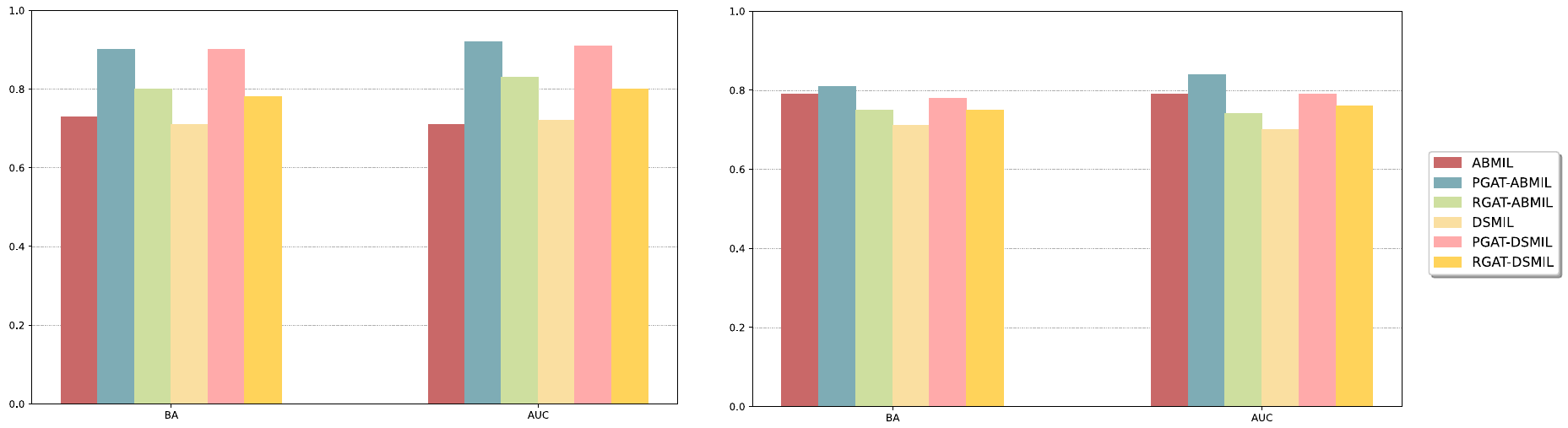}}
  \caption{Performance Comparison of Baseline and GAT-based MIL Models Across Camelyon16 (left) and Camelyon17 (right) Datasets.}
  \label{fig:spatialcontext}
\end{figure*}

\section{Results}
\subsection{Impact of Graph Construction}


The experimental results presented in Table \ref{tab:graphs_representation} demonstrate that the Patch-Graphs consistently outperform other graph representations across all evaluated metrics, with an improvement of 16$\%$ over the highest recall obtained by Local Region-GCN and a 20$\%$ improvement over the best accuracy and AUC achieved by Centroid-GCN.

These findings suggest a strong correlation  between graph representation compactness and performance. As the representation becomes more compact, performance tends to decrease across all metrics.
Patch-Graphs have superior performance, as indicated by the highest accuracy, recall, and AUC scores, due to their ability to preserve the original information of WSI. In contrast, Region-Graphs generate a more compact representation by having each node represent a region, a set of patches, resulting in lower performance metrics. The Centroid-Graphs, which drastically reduce the representation to only nine nodes, show the lowest performance, highlighting the trade-off between preserving critical information and performance.

When comparing the two RAG strategies, we find that, contrary to the hypothesis that dataset-level clustering would offer a broader understanding of tissue types and improve performance, the results show otherwise. Clustering at the image level provides more adaptability, as it better captures the unique features and patterns within each image.

\begin{table*}[t]
 \caption{Results on Camelyon16 and Camelyon17 datasets, with and without Interventional Training. The $\Delta$ values indicate the performance difference introduced by Interventional Training. \textcolor{red}{Red} indicates a decrease in performance, while \textcolor{green}{green} denotes an improvement.}
 \vspace{0.2cm}
 \centering
 \scriptsize{
 \renewcommand{\arraystretch}{1.4}
    \resizebox{\textwidth}{!}{
    \begin{tabular}{ c *{4}{c} | *{4}{c}}
    \hline
     & \multicolumn{4}{c}{  \textbf{Camelyon 16}} & \multicolumn{4}{c}{  \textbf{Camelyon 17}}\\ \cline{2-9}
     \textbf{Configuration} & BA & AUC & F1 & Precision & BA & AUC & F1 & Precision \\ \hline
     \multicolumn{1}{c}{\textbf{ABMIL}} &  0.726\ensuremath{\pm}0.17 &  0.711\ensuremath{\pm}0.21 &  0.713\ensuremath{\pm}0.20 &  0.745\ensuremath{\pm}0.18 &  0.788\ensuremath{\pm}0.06 &  0.790\ensuremath{\pm}0.08 &  0.803\ensuremath{\pm}0.07 &  0.838\ensuremath{\pm}0.08\\ 
     \multicolumn{1}{c}{\textbf{w\textbackslash IT}} &  0.894\ensuremath{\pm}0.05 &  0.920\ensuremath{\pm}0.04 &  0.896\ensuremath{\pm}0.05 &  0.901\ensuremath{\pm}0.04 &  0.820\ensuremath{\pm}0.04 &  0.852\ensuremath{\pm}0.03  &  0.830\ensuremath{\pm}0.04 &  0.855\ensuremath{\pm}0.04 \\ 
     \multicolumn{1}{c}{\textbf{$\Delta$}} & \textcolor{green}{0.168} & \textcolor{green}{0.209} & \textcolor{green}{0.183} & \textcolor{green}{0.156} & \textcolor{green}{0.032} & \textcolor{green}{0.062} & \textcolor{green}{0.027} & \textcolor{green}{0.017}\\ \hline
     \multicolumn{1}{c}{\textbf{DSMIL}} &  0.713\ensuremath{\pm}0.04 &  0.724\ensuremath{\pm}0.06 &  0.707\ensuremath{\pm}0.04 &  0.726\ensuremath{\pm}0.05 &  0.708\ensuremath{\pm}0.07 &  0.703\ensuremath{\pm}0.08 &  0.706\ensuremath{\pm}0.10 &  0.830\ensuremath{\pm}0.03\\ 
     \multicolumn{1}{c}{\textbf{w\textbackslash IT}} &  0.764\ensuremath{\pm}0.10 &  0.781\ensuremath{\pm}0.10 &  0.759\ensuremath{\pm}0.11 &  0.771\ensuremath{\pm}0.10 &  0.746\ensuremath{\pm}0.07 &  0.768\ensuremath{\pm}0.06  &  0.708\ensuremath{\pm}0.08 &  0.743\ensuremath{\pm}0.08 \\ 
     \multicolumn{1}{c}{\textbf{$\Delta$}} & \textcolor{green}{0.051} & \textcolor{green}{0.057} & \textcolor{green}{0.052} & \textcolor{green}{0.045} & \textcolor{green}{0.038} & \textcolor{green}{0.065} & \textcolor{green}{0.002} & \textcolor{red}{0.087}\\ \hline
    \multicolumn{1}{c}{\textbf{PatchGAT-ABMIL}} & 0.901\ensuremath{\pm}0.06 & 0.923\ensuremath{\pm}0.06 &  0.899\ensuremath{\pm}0.06 & 0.904\ensuremath{\pm}0.06 & 0.814\ensuremath{\pm}0.04& 0.837\ensuremath{\pm}0.04 &  0.821\ensuremath{\pm}0.04 & 0.850\ensuremath{\pm}0.05\\ 
    \multicolumn{1}{c}{\textbf{w\textbackslash IT}} & 0.892\ensuremath{\pm}0.03 & 0.910\ensuremath{\pm}0.04 &  0.889\ensuremath{\pm}0.05 &  0.895\ensuremath{\pm}0.05 & 0.811\ensuremath{\pm}0.05 &  0.823\ensuremath{\pm}0.07  & 0.812\ensuremath{\pm}0.07 &  0.822\ensuremath{\pm}0.08 \\ 
    \multicolumn{1}{c}{\textbf{$\Delta$}} & \textcolor{red}{0.009} & \textcolor{red}{0.013} & \textcolor{red}{0.010} & \textcolor{red}{0.009} & \textcolor{red}{0.002} & \textcolor{red}{0.014} & \textcolor{red}{0.009} & \textcolor{red}{0.029}\\ \hline   \multicolumn{1}{c}{\textbf{PatchGAT-DSMIL}} & 0.888\ensuremath{\pm}0.06 & 0.903\ensuremath{\pm}0.07 &  0.885\ensuremath{\pm}0.07 &  0.887\ensuremath{\pm}0.07 &  0.778\ensuremath{\pm}0.07&  0.787\ensuremath{\pm}0.07 &  0.795\ensuremath{\pm}0.07 & 0.861\ensuremath{\pm}0.05\\ 
    \multicolumn{1}{c}{\textbf{w\textbackslash IT}} & 0.866\ensuremath{\pm}0.06 & 0.887\ensuremath{\pm}0.06 &  0.875\ensuremath{\pm}0.06 &  0.896\ensuremath{\pm}0.05 & 0.746\ensuremath{\pm}0.09 & 0.748\ensuremath{\pm}0.12 & 0.755\ensuremath{\pm}0.11 & 0.787\ensuremath{\pm}0.12\\ 
    \multicolumn{1}{c}{\textbf{$\Delta$}} & \textcolor{red}{0.022} & \textcolor{red}{0.016} & \textcolor{red}{0.010} & \textcolor{green}{0.009} & \textcolor{red}{0.033} & \textcolor{red}{0.039} & \textcolor{red}{0.040} & \textcolor{red}{0.073}\\ \hline
   \multicolumn{1}{c}{\textbf{RegionGAT-ABMIL}} &  0.798\ensuremath{\pm}0.04 & 0.827\ensuremath{\pm}0.05 & 0.793\ensuremath{\pm}0.04 &  0.794\ensuremath{\pm}0.04 & 0.752\ensuremath{\pm}0.04 & 0.742\ensuremath{\pm}0.04 &0.747\ensuremath{\pm}0.05 & 0.753\ensuremath{\pm}0.07\\ 
    \multicolumn{1}{c}{\textbf{w\textbackslash IT}} &  0.783\ensuremath{\pm}0.06 & 0.811\ensuremath{\pm}0.07 &  0.785\ensuremath{\pm}0.07&  0.794\ensuremath{\pm}0.06 & 0.713\ensuremath{\pm}0.08 &  0.706\ensuremath{\pm}0.1&  0.741\ensuremath{\pm}0.13 &  0.773\ensuremath{\pm}0.12\\ 
    \multicolumn{1}{c}{\textbf{$\Delta$}} & \textcolor{red}{0.015} & \textcolor{red}{0.015} & \textcolor{red}{0.008} & \textcolor{gray}{0.000} & \textcolor{red}{0.039} & \textcolor{red}{0.037} & \textcolor{red}{0.006} & \textcolor{green}{0.020}\\ \hline
    \multicolumn{1}{c}{\textbf{RegionGAT-DSMIL}} & 0.787\ensuremath{\pm}0.02 & 0.788\ensuremath{\pm}0.04 & 0.768\ensuremath{\pm}0.08 &  0.776\ensuremath{\pm}0.06 & 0.769\ensuremath{\pm}0.07 &  0.772\ensuremath{\pm}0.09 &  0.764\ensuremath{\pm}0.10 & 0.809\ensuremath{\pm}0.08\\ 
    \multicolumn{1}{c}{\textbf{w\textbackslash IT}} & 0.754\ensuremath{\pm}0.08 &  0.760\ensuremath{\pm}0.10 &  0.751\ensuremath{\pm}0.10 &  0.793\ensuremath{\pm}0.08 & 0.701\ensuremath{\pm}0.09 & 0.722\ensuremath{\pm}0.12 & 0.727\ensuremath{\pm}0.10 &  0.775\ensuremath{\pm}0.12\\ 
    \multicolumn{1}{c}{\textbf{$\Delta$}} & \textcolor{red}{0.033} & \textcolor{red}{0.028} & \textcolor{red}{0.017} & \textcolor{green}{0.017} & \textcolor{red}{0.068} & \textcolor{red}{0.05} & \textcolor{red}{0.037} & \textcolor{red}{0.034}\\
    \hline
    \end{tabular}
    }
}
\label{tab:InterventionResults}
\end{table*}

\subsection{Impact of Spatial Context in WSI Analysis}
\label{subsec:spatial_context_results}

As shown in figure \ref{fig:spatialcontext}, the proposed method achieves higher metrics compared to baseline models (ABMIL and DSMIL) on both datasets. PatchGAT-ABMIL outperforms all approaches, with 17 \% improvement in balanced accuracy and 21 \% in AUC over ABMIL. This shows that incorporating spatial information is crucial in the context of metastases detection, because it helps the model better understand the underlying structure of the tissues, resulting in improved prediction accuracy. 
The decrease in performance for Camelyon17, when compared to Camelyon16 is expected. In Camelyon17, the training and testing were designed to reinforce the distribution shifts present in the dataset, as it includes data from multiple medical centers and different scanners. This setup introduces significant variability between the training and testing sets, making the task more challenging. In contrast, although Camelyon16 data also comes from two different medical centers, the distribution shift was not explicitly enforced during training and testing. As a result, the training and testing sets for each fold in Camelyon16 are more similar, leading to higher performance metrics. Despite the explicit domain shift in Camelyon17, where each test set in each fold contains data from a single medical center, graph-based models still demonstrate better generalization compared to traditional MIL models.





\subsection{Impact of Interventions}
\label{subsec:interventions_resultssec}

Table \ref{tab:InterventionResults} outlines the results of incorporating interventions into a Graph-based Multiple Instance Learning (MIL) pipeline. While it was initially hypothesized that these interventions would enhance model robustness and generalization, the results show a consistent decline in performance across all evaluated metrics and datasets for the graph-based approaches. Additionally, we observe that the standard deviations (std) for MIL w{\textbackslash}IT are generally larger compared to those of the graph-based models. Moreover, the performance results of the graph-based models tend to fall within the range of those obtained with MIL, further emphasizing the stability of graph-based approaches compared to MIL w{\textbackslash}IT.

It is crucial to explore why these interventions may have negatively impacted the learning in graph-based models, such as GATs. These models rely on structured representations that naturally filter out irrelevant information and focus on critical, discriminative features. This inherent filtering likely contributes to their more stable performance and reduced variability \citep{velickovic2018}. To investigate this further, we analyze the construction of the confounder dictionary and visualize the t-SNE plot of the bag embeddings (please refer to Appendix \ref{subsec:bag_embapp}). The distinct separation of clusters suggests that graph-based models are capable of pushing feature representations of different classes further apart. This indicates that GAT-ABMIL shifts the model’s focus from potentially visual patterns to more discriminative features, enhancing class separation, which could explain why interventions result in a decrease in performance. The effectiveness of the graph-based model is further highlighted by the cluster purity analysis for GAT-ABMIL, with a score of 1.0, which indicates that each cluster contains only instances from a single class. Detailed examination shows that centroid 0 corresponds to the `normal' class, while centroid 1 represents the `tumor' class.

This aligns with the findings of \citep{Guo2024}, who suggest that the self-attention mechanism in Graph Attention Networks follows the information bottleneck principle, which may explain the clear separation between clusters, since it is filtering out irrelevant information and retaining only the relevant features. As \citep{Yang2024IB} points out, information bottleneck enables GNNs to learn invariant features, effectively reducing the impact of confounding factors, which might explain the reason why the performance drops when introducing interventions.



\subsection{Qualitative Analysis}
We perform a qualitative analysis by illustrating attention heatmaps in figure \ref{fig:all_heatmaps} in appendix (\ref{app:qa}). These heatmaps are an interpretability visual tool to corroborate the classification results, comparing the baseline ABMIL model with GMIL both without and with interventional training. Through visual inspection, we can reinforce the importance of considering the spatial relationship between patches for WSI classification. The introduction of the interventions lead to less focused attention, as evidenced by the comparison between the last two columns. This suggests that interventions may disrupt the model's ability to capture relevant spatial information, reducing the classification performance. Our findings highlight the importance of preserving spatial context.

\section{Conclusion}

This work proposes a novel evaluation pipeline designed to assess the robustness of WSI classification models. We conduct a thorough comparison of multiple graph construction techniques, MIL models, graph-MIL approaches, and interventional training. We also introduce a new framework, GMIL-IT, to understand the impact of each component in model generalization under domain shifts. Our analysis showed that graph-based models alone outperform models enhanced with interventional training, highlighting the robustness of the graph structures.

\section{Limitations}

We acknowledge that the datasets used in this thesis do not demonstrate the models’ ability to generalize to other types of cancer. Our primary goal was to analyze the robustness of graph-based
models on real-world datasets containing domain shifts. The Camelyon17 dataset, with data from five different hospitals and three scanners, provided an ideal setting to evaluate generalization performance
under these shifts, which, to the best of our knowledge, is a novel way of assessing WSI classifiers using this dataset. This approach emphasizes the strength of the proposed method in handling real-
world variability. To further validate our findings, we extended the analysis to another publicly available dataset for clear cell renal cell carcinoma, MMIST-ccRCC \citep{Mota_2024_CVPR}, which includes data from two sources. The
task here is 12-month survival prediction, and results indicate that graph-based approaches outperform traditional methods (ABMIL 64.9\% balanced accuracy vs. PatchGAT-ABMIL 71.1\%). These additional results reinforce the original claims and highlight the robustness of graph based models across different datasets.

However, to fully confirm the generalizability of these models, future work should validate them on additional datasets covering a broader range of cancer types. 
Furthermore, the findings show that the backdoor adjustment intervention did not improve the performance of graph-based models in our experiments, and that it might be worth exploring alternative strategies could clarify whether graphs are inherently robust to any type of intervention.

\label{sec:cite}
\bibliography{jmlr-sample}

\appendix

\section{Detailed Results}\label{apd:first}

\subsection{Additional Results of Section \ref{subsec:spatial_context_results}}
\begin{figure*}[htbp]
  {\includegraphics[width=1.0\linewidth]{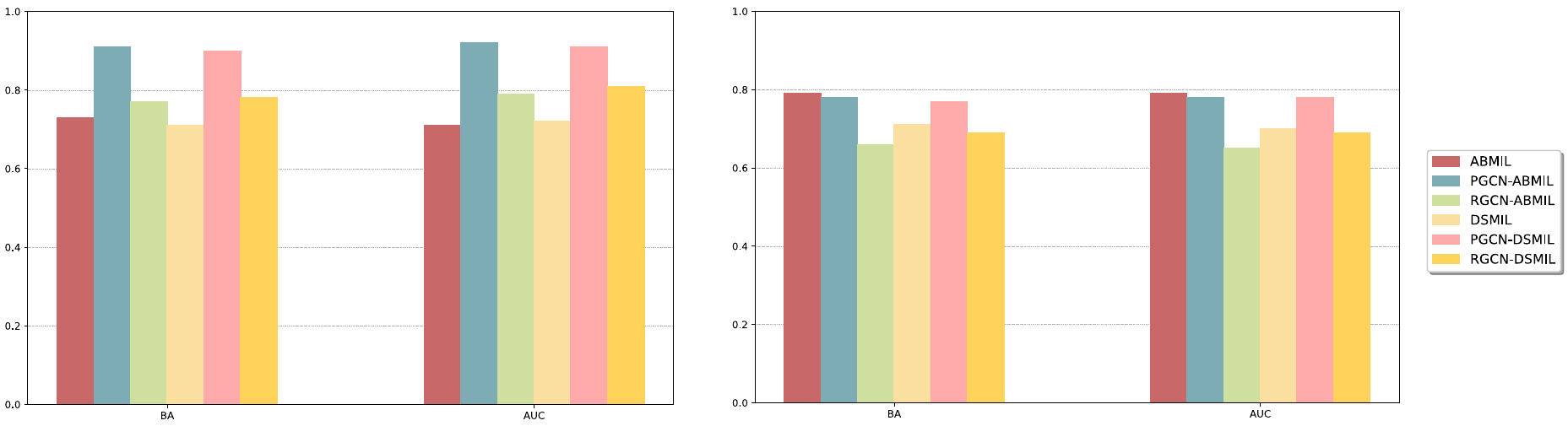}}
  \caption{Performance Comparison of Baseline and GCN-based MIL Models Across Camelyon16 (left) and Camelyon17 (right) Datasets.}
  \label{fig:spatialcontextgcn}
\end{figure*}

Figure \ref{fig:spatialcontextgcn} illustrates the additional results obtained for GCN-based model for both datasets. 

\subsection{Additional Results of Section \ref{subsec:interventions_resultssec}}
Table \ref{tab:interventions_gcn} outlines the results of the experiments with Interventional Training for GCN-based model.

\begin{table*}[!htbp]
 \caption{Results on Camelyon16 and Camelyon17 datasets, with and without Interventional Training. The $\Delta$ values indicate the performance difference introduced by Interventional Training. \textcolor{red}{Red} indicates a decrease in performance, while \textcolor{green}{green} denotes an improvement.}
 \vspace{0.2cm}
 \centering
 \scriptsize{
 \renewcommand{\arraystretch}{1.4}
    \resizebox{\textwidth}{!}{
    \begin{tabular}{ c *{4}{c} | *{4}{c}}
    \hline
     & \multicolumn{4}{c}{  \textbf{Camelyon 16}} & \multicolumn{4}{c}{  \textbf{Camelyon 17}}\\ \cline{2-9}
     \textbf{Configuration} & BA & AUC & F1 & Precision & BA & AUC & F1 & Precision \\ \hline
     \multicolumn{1}{c}{\textbf{ABMIL}} &  0.726\ensuremath{\pm}0.17 &  0.711\ensuremath{\pm}0.21 &  0.713\ensuremath{\pm}0.20 &  0.745\ensuremath{\pm}0.18 &  0.788\ensuremath{\pm}0.06 &  0.790\ensuremath{\pm}0.08 &  0.803\ensuremath{\pm}0.07 &  0.838\ensuremath{\pm}0.08\\ 
     \multicolumn{1}{c}{\textbf{w\textbackslash IT}} &  0.894\ensuremath{\pm}0.05 &  0.920\ensuremath{\pm}0.04 &  0.896\ensuremath{\pm}0.05 &  0.901\ensuremath{\pm}0.04 &  0.820\ensuremath{\pm}0.04 &  0.852\ensuremath{\pm}0.03  &  0.830\ensuremath{\pm}0.04 &  0.855\ensuremath{\pm}0.04 \\ 
     \multicolumn{1}{c}{\textbf{$\Delta$}} & \text{\textcolor{green}{0.168}} & \text{\textcolor{green}{0.209}} & \text{\textcolor{green}{0.183}} & \text{\textcolor{green}{0.156}} & \text{\textcolor{green}{0.032}} & \text{\textcolor{green}{0.062}} & \text{\textcolor{green}{0.027}} & \text{\textcolor{green}{0.017}}\\ \hline
     \multicolumn{1}{c}{\textbf{DSMIL}} &  0.713\ensuremath{\pm}0.04 &  0.724\ensuremath{\pm}0.06 &  0.707\ensuremath{\pm}0.04 &  0.726\ensuremath{\pm}0.05 &  0.708\ensuremath{\pm}0.07 &  0.703\ensuremath{\pm}0.08 &  0.706\ensuremath{\pm}0.10 &  0.830\ensuremath{\pm}0.03\\ 
     \multicolumn{1}{c}{\textbf{w\textbackslash IT}} &  0.764\ensuremath{\pm}0.10 &  0.781\ensuremath{\pm}0.10 &  0.759\ensuremath{\pm}0.11 &  0.771\ensuremath{\pm}0.10 &  0.746\ensuremath{\pm}0.07 &  0.768\ensuremath{\pm}0.06  &  0.708\ensuremath{\pm}0.08 &  0.743\ensuremath{\pm}0.08 \\ 
     \multicolumn{1}{c}{\textbf{$\Delta$}} & \text{\textcolor{green}{0.051}} & \text{\textcolor{green}{0.057}} & \text{\textcolor{green}{0.052}} & \text{\textcolor{green}{0.045}} & \text{\textcolor{green}{0.038}} & \text{\textcolor{green}{0.065}} & \text{\textcolor{green}{0.002}} & \text{\textcolor{red}{0.087}}\\ \hline
    \multicolumn{1}{c}{\textbf{PatchGCN-ABMIL}} & 0.906\ensuremath{\pm}0.04 & 0.916\ensuremath{\pm}0.05 &  0.907\ensuremath{\pm}0.04 & 0.911\ensuremath{\pm}0.04 & 0.777\ensuremath{\pm}0.06& 0.783\ensuremath{\pm}0.05 &  0.776\ensuremath{\pm}0.09 & 0.803\ensuremath{\pm}0.11\\ 
    \multicolumn{1}{c}{\textbf{w\textbackslash IT}} & 0.890\ensuremath{\pm}0.06 & 0.887\ensuremath{\pm}0.07 &  0.894\ensuremath{\pm}0.06 &  0.904\ensuremath{\pm}0.04 & 0.787\ensuremath{\pm}0.07 &  0.799\ensuremath{\pm}0.07  & 0.818\ensuremath{\pm}0.08 &  0.840\ensuremath{\pm}0.08 \\ 
    \multicolumn{1}{c}{\textbf{$\Delta$}} & \text{\textcolor{red}{0.016}} & \text{\textcolor{red}{0.029}} & \text{\textcolor{red}{0.013}} & \text{\textcolor{red}{0.007}} & \text{\textcolor{green}{0.010}} & \text{\textcolor{green}{0.015}} & \text{\textcolor{green}{0.042}} & \text{\textcolor{red}{0.037}}\\ \hline   
    \multicolumn{1}{c}{\textbf{PatchGCN-DSMIL}} & 0.896\ensuremath{\pm}0.03 & 0.915\ensuremath{\pm}0.05 &  0.897\ensuremath{\pm}0.03 &  0.906\ensuremath{\pm}0.03 &  0.772\ensuremath{\pm}0.08&  0.775\ensuremath{\pm}0.08 &  0.778\ensuremath{\pm}0.09 & 0.797\ensuremath{\pm}0.11\\ 
    \multicolumn{1}{c}{\textbf{w\textbackslash IT}} & 0.860\ensuremath{\pm}0.05 & 0.898\ensuremath{\pm}0.07 &  0.862\ensuremath{\pm}0.06 &  0.866\ensuremath{\pm}0.06 & 0.771\ensuremath{\pm}0.08 & 0.778\ensuremath{\pm}0.07 & 0.779\ensuremath{\pm}0.10 & 0.807\ensuremath{\pm}0.12\\ 
    \multicolumn{1}{c}{\textbf{$\Delta$}} & \text{\textcolor{red}{0.036}} & \text{\textcolor{red}{0.017}} & \text{\textcolor{red}{0.035}} & \text{\textcolor{red}{0.039}} & \text{\textcolor{red}{0.001}} & \text{\textcolor{green}{0.002}} & \text{\textcolor{green}{0.001}} & \text{\textcolor{green}{0.010}}\\ \hline
   \multicolumn{1}{c}{\textbf{RegionGCN-ABMIL}} &  0.768\ensuremath{\pm}0.03 & 0.781\ensuremath{\pm}0.05 & 0.774\ensuremath{\pm}0.06 &  0.798\ensuremath{\pm}0.06 & 0.754\ensuremath{\pm}0.05 & 0.788\ensuremath{\pm}0.05 & 0.744\ensuremath{\pm}0.08 &  0.749\ensuremath{\pm}0.12 \\ 
    \multicolumn{1}{c}{\textbf{w\textbackslash IT}} &  0.781\ensuremath{\pm}0.06 & 0.801\ensuremath{\pm}0.04 & 0.793\ensuremath{\pm}0.07 & 0.798\ensuremath{\pm}0.07 &  0.742\ensuremath{\pm}0.06 & 0.747\ensuremath{\pm}0.05 &  0.721\ensuremath{\pm}0.09 & 0.762\ensuremath{\pm}0.09\\ 
    \multicolumn{1}{c}{\textbf{$\Delta$}} & \text{\textcolor{green}{0.028}} & \text{\textcolor{green}{0.006}} & \text{\textcolor{green}{0.012}} & \text{\textcolor{red}{0.000}} & \text{\textcolor{red}{0.012}} & \text{\textcolor{red}{0.041}} & \text{\textcolor{red}{0.023}} & \text{\textcolor{green}{0.013}}\\ \hline
    \end{tabular}}}
    \label{tab:interventions_gcn}
\end{table*}

\subsection{Visualization of Bag Embeddings}
\label{subsec:bag_embapp}
Figures \ref{fig:tsne} and \ref{fig:tsne_c17} illustrate the t-SNE representation of the bag embeddings of the training set for Camelyon16 and Camelyon17, respectively. In the case of GAT-ABMIL the graph representation chosen was patch graphs. In Camelyon17, we include the t-SNE plots for multiple folds,  highlighting the medical centers and scanners for each fold. Our goal is to analyze how the domain shifts affect the data distribution and how it impacts the performance of the model. 
\begin{samepage}
\begin{figure*}[!htbp]

\floatconts
  {fig:tsne} 
  {\caption{t-SNE visualization of bag embeddings from Camelyon16, comparing ABMIL (left) and GAT-ABMIL (right) for confounder dictionary construction.}} 
  { 
    {
      \includegraphics[width=0.38\linewidth]{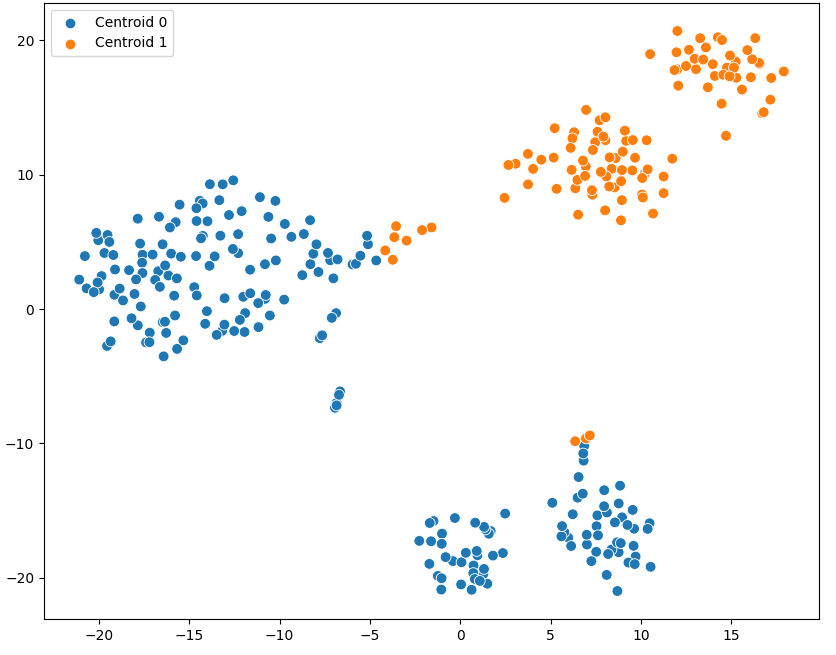}
      \label{fig:abmil_tsne}
    }
    \qquad
    {
      \includegraphics[width=0.38\linewidth]{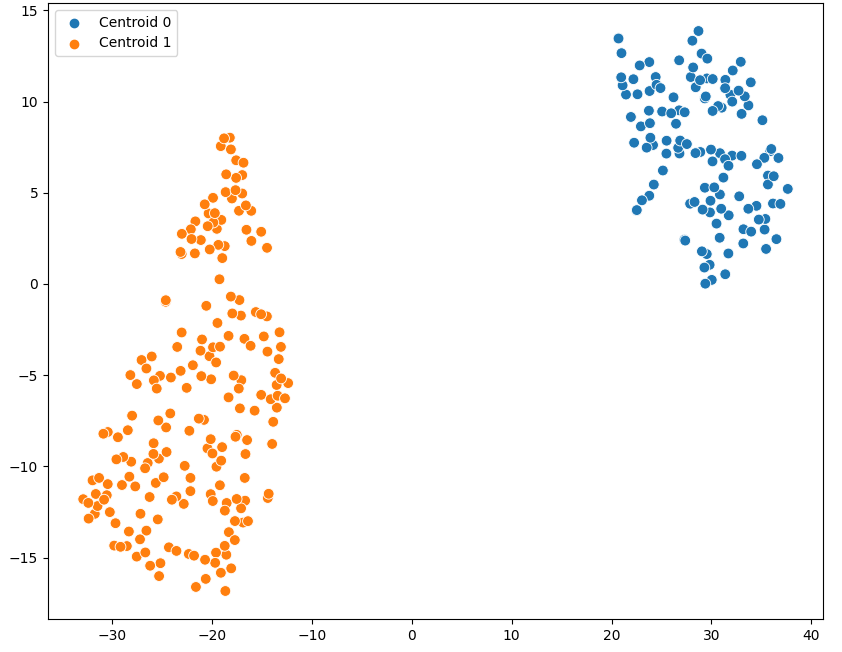}
      \label{fig:gatabmil_tsne}
    }
  }

\end{figure*}

\begin{figure*}[!htbp]
  \floatconts
  {fig:tsne_c17} 
  {\caption{t-SNE visualization of bag embeddings from Camelyon17, comparing GAT-ABMIL fold 0 (left) and GAT-ABMIL fold 2 (right) for confounder dictionary construction.}} 
  { 
    {
      \includegraphics[width=0.46\linewidth]{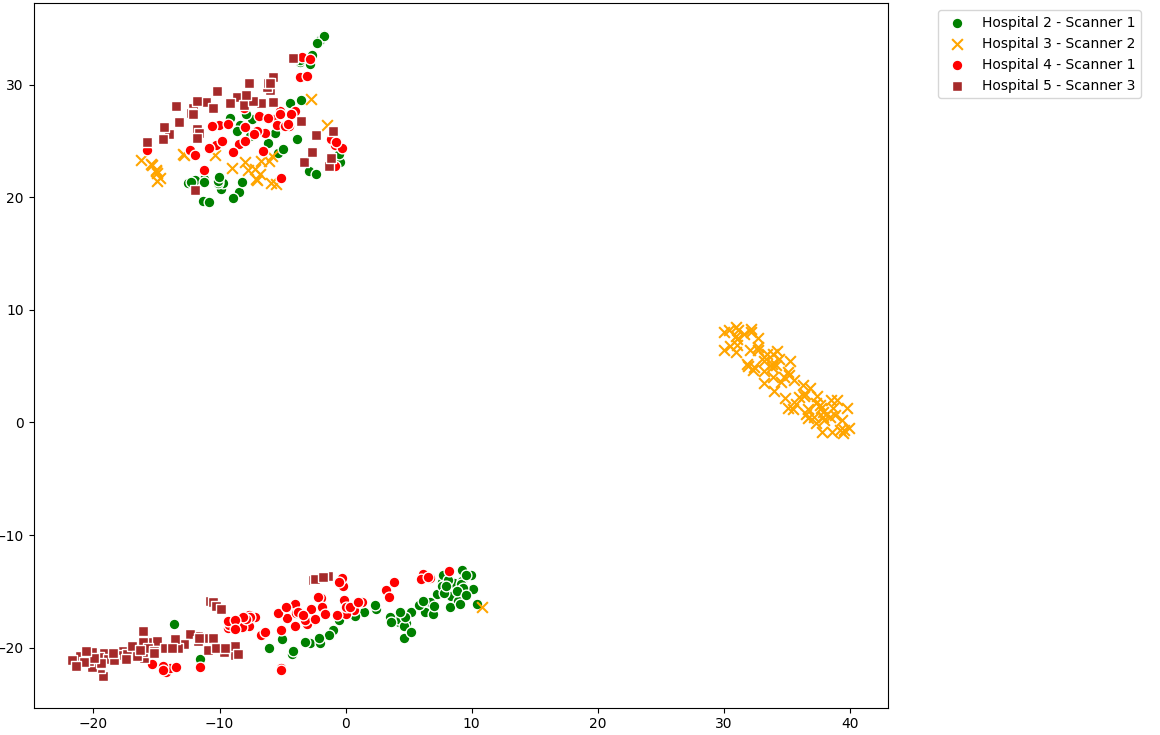}
      \label{fig:fold0_tsne_c17}
    }
    \qquad
    {
      \includegraphics[width=0.46\linewidth]{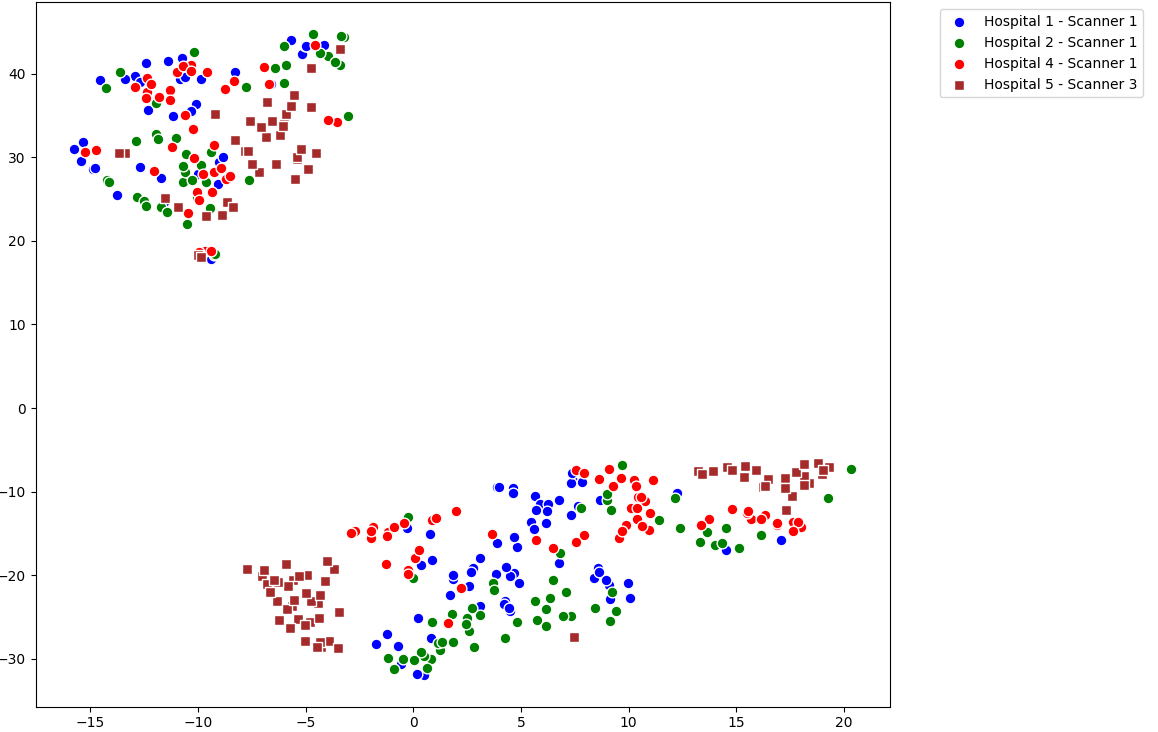}
      \label{fig:fold2_tsne_c17}
    }
  }

\end{figure*}
\end{samepage}

\subsection{Qualitative Analysis} \label{app:qa}
\begin{figure*}[htbp]
  {\includegraphics[width=0.87\linewidth]{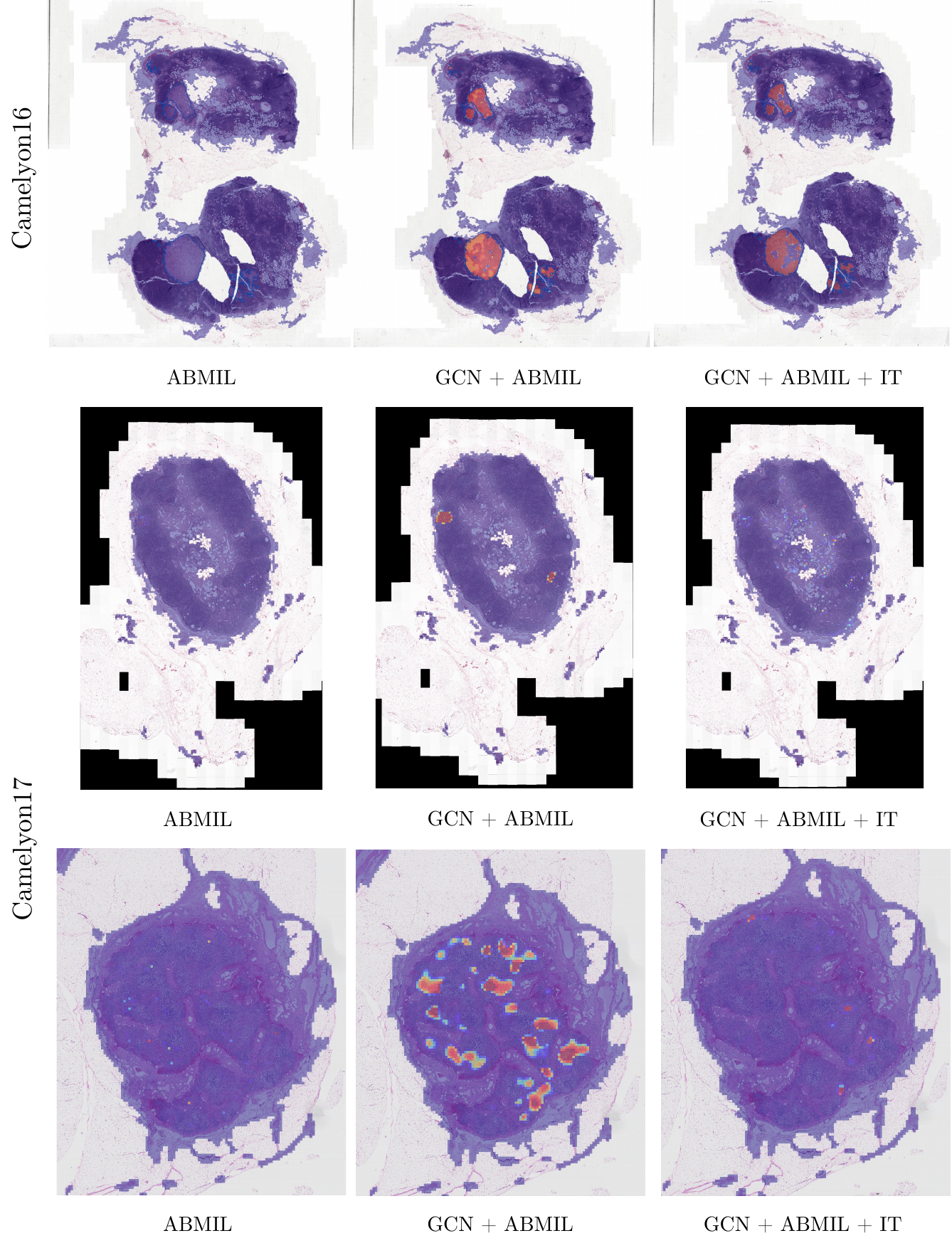}}
  \caption{Heatmap visualizations for a tumor slides of Camelyon16 and Camelyon17. The blue lines outline tumor regions. \textcolor{red}{Red} color indicates patches with high attention score, while \textcolor{blue}{blue} indicate patches with lowest attention score.}
  \label{fig:all_heatmaps}
\end{figure*}

Figure \ref{fig:all_heatmaps} illustrates the attention heatmaps obtained ABMIL model with GMIL both without and with interventional training.

\end{document}